# Network Abstractions of Prescription Patterns in a Medicaid Population

**Radhakrishnan Nagarajan, Ph.D.[1*], Jeffery Talbert, Ph.D.[2]**
**[1]Division of Biomedical Informatics, [2]Department of Pharmacy Practice and Science,**
**University of Kentucky, KY, USA**

**Abstract**

*Understanding prescription patterns have relied largely on aggregate statistical measures. Evidence of doctor-shopping, inappropriate prescribing, drug diversion and patient seeking prescription drugs across multiple prescribers demand understanding the concerted working of prescribers and prescriber communities as opposed to treating them as independent entities. We model potential associations between prescribers as prescriber-prescriber network (PPN) and subsequently investigate its properties across Schedule II, III, IV drugs in a single month in a Medicaid population. Community structure detection algorithms and geo-spatial layouts revealed characteristic patterns in PPN markedly different from their random graph surrogate counterparts rejecting them as potential generative mechanism. Outlier detection with recommended thresholds also revealed a subset of prescriber specialties to be constitutively flagged across Schedule II, III, IV drugs. Presence of prescriber communities may assist in targeted monitoring and their deviation from random graphs may serve as a metric in assessing PPN evolution temporally and pre-/post- interventions.*

**Introduction**

The Centers for Disease Control (CDC) identifies prescription drug abuse as an epidemic and a leading preventable cause of death in the United States (U.S.)[1,2]. In 2013, 22,000 drug related deaths from prescription drugs were about one-half of all drug related deaths[1]. Among prescription drugs, the abuse of prescription opioids led to more than 16,000 deaths each year, about one death every 35 minutes, more than the number of deaths linked to the use of heroin and cocaine combined[3]. The total number of opioid prescriptions dispensed from retail pharmacies tripled between 1990 and 2010, exceeding 210 million by 2010[4]. Over 12 million people reported abuse of prescription opioids in 2010 resulting in more than 475,000 emergency department visits (National Survey on Drug Use and Health, NSDUH 2010; Drug Abuse Warning Network, DAWN 2010). During the 1990s, development of evidence-based guidelines on pain management, and aggressive pharmaceutical marketing converged to initiate a shift in the use of opioids for pain management[5,6]. Excluding marijuana, the prevalence of prescription drug abuse exceeds that for all illicit drugs combined (Substance Abuse and Mental Health Services Administration, SAMHSA 2006). The focus of the present study is on Kentucky, a state ranked 2nd nationally in prescription drug overdose mortality (24.6 deaths per 100,000), ranked 3rd nationally in utilization of prescription opioids (128 prescriptions per 100 people) and that has seen an increasing prescription drug abuse problem for the past 10 years. The state was an early adopter of electronic prescription drug monitoring programs, through the Kentucky All Schedule Prescription Electronic Reporting (KASPER) program, but prescription drug abuse and diversion remains a persistent public health issue.

Traditional approaches for understanding prescription patterns have focused primarily on aggregate measures such as prescription counts (i.e. number of prescriptions) and their temporal trends across prescribers as well as drugs[7-10]. While useful, they implicitly subscribe to reductionism and the where prescribers are treated as independent entities, **Fig. 1**. However, there is growing evidence of doctor-shopping, inappropriate prescribing and drug diversion[10-13] with patients seeking drugs across multiple prescribers establishing an indirect association between them. The present study uses network abstractions to visualize and investigate the concerted working of prescribers as a system (prescriber-prescriber network, **PPN**), **Fig. 1**. Subsequently, it investigates the presence of prescriber communities in PPN and the usefulness of random graph models as potential generative mechanism of the observed patterns in PPN across three scheduled drugs (Schedule II, III, IV). Geo-spatial layout of the PPN provides a convenient visualization of the patient movement between prescribers across the state. Presence of community structures has the potential to assist in targeted monitoring and surveillance of select prescriber communities. Quantifying potential differences between the community structures in PPN from that of random graph models (controls) may serve as a useful metric in understanding the temporal evolution of PPN topology as well as changes in PPN topology to interventions in a pre-/post- setting. Previous studies have successfully used social network analysis for understanding social influences on substance abuse[14]. Such an understanding has also been shown to be useful in implementing effective intervention and prevention programs as well as monitoring and evaluation[15,16]. Studies have also investigated the co-prescription networks and concluded these networks to exhibit characteristics such as scale invariance[17]. A recent study reported correlation between network metrics and extreme prescription patterns in hospitals[18] while others have demonstrated the advantages of networks for visualizing large healthcare data and their usefulness from cognition and hypothesis generation standpoints[19].

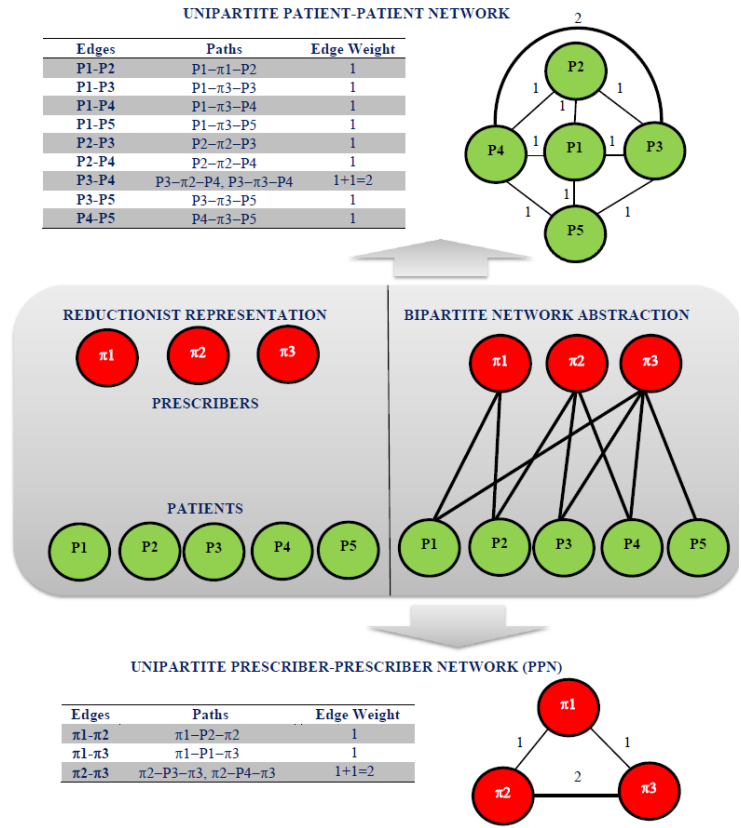

**Figure 1.** Classical reductionist representation (middle, left) of the prescribers (red circles, $\pi 1, \pi 2, \pi 3$) and the patients (green circles, P1, P2, P3, P4, P5) as independent entities and bipartite network abstraction of patients and prescribers as a result of patient seeking drugs across multiple prescribers is also shown (middle, right). Construction of the unipartite projections from the bipartite network abstraction resulting patient-patient network (top) and prescriber-prescriber network (PPN, bottom) represented as weighted undirected graphs are also shown.

## Materials and Methods

*Data Description*: Prescription datasets were retrieved from Kentucky Medicaid Services and consisted of encrypted prescriber IDs (National Provider Identifier), encrypted patient IDs, and GIS locations of the prescribers and their specialty information in the month of September 2011 across a spectrum of controlled substances drugs including Schedule II, III and IV drugs. The University of Kentucky Institutional Review Board approved the study prior to obtaining the data sets. The different schedules and typical drugs associated with each schedule are shown in **Table 1**. Drugs under each of the schedule from the data in the present study are shown in **Table 2**. The data investigated are outpatient pharmaceutical claims where prescribers included individual physicians, nurse practitioners, and dentists. The data does not include prescriptions dispensed in inpatient settings such as hospitals. The data sets were restricted to a single month (30 day period) to focus on a snapshot of typical prescribing patterns in a controlled time frame with minimal likelihood of a patient seeking drugs of the same schedule across multiple prescribers. The number of prescriptions dispensed across Schedule II, III, IV were 15,108, 45,692 and 60,578 respectively.

**Table 1.** Scheduled Drugs: Definitions and Examples

| Scheduled Drugs | Definition | Examples |
|---|---|---|
| Schedule II | Has an accepted medical use in treatment in United States but has a high abuse potential. Abuse may lead to severe psychological or physical dependence | Morphine, Cocaine, Oxycodone, Hydrocodone*, Barbiturates, Amphetamines |
| Schedule III | Has a potential for abuse less than the drugs in schedules I and II. Has a currently accepted medical use in treatment in the United States. Abuse may lead to moderate or low physical dependence or high psychological dependence | Codeine, Ketamine, Anabolic steroids, Suboxone |
| Schedule IV | Has a low potential for abuse relative to the drugs in schedule III. Has a currently accepted medical use in treatment in the United States. Abuse may lead to limited physical or psychological dependence relative to the drugs in Schedule III | Diazepam, Alprazolam, Phenobarbital, Lorazepam, Zolpidem, Phentermine |
| Schedule V | Has a low potential for abuse relative to the drugs in schedule IV. Has a currently accepted medical use in treatment in the United States. Abuse may lead to limited physical or psychological dependence relative to the drugs in Schedule IV | OTC Cough syrups containing codeine, Diphenoxylate/atropine |

* For the duration of this study, Hydrocodone was classified as Schedule III. However, it was reclassified from Schedule III to Schedule II in 2014

*Prescribers with Questionable Prescription Patterns*: A recent report from the Department of Health and Human Services, Office of Inspector General[20] had investigated five critical measures for identifying prescribers with questionable prescription patterns. More specifically, Tukey's outlier detection approach in conjunction with conservative thresholds was used to flag prescribers. Since the patient distribution across the prescribers can be non-uniform, the present study investigated the univariate distribution of the average prescription counts as opposed to the total prescription counts of the prescribers across each of the Schedule II, III, IV drugs. Average prescription count of a prescriber was determined as the ratio of the total number of prescriptions written by a prescriber by the total number of distinct patients who obtained prescriptions from that specific prescriber. Using recommended and conservative thresholds[20], prescribers were flagged as "extreme outliers" if their prescription counts were greater than the $Q_3 + 4.5$ IQR, where $Q_3$ represents the $3^{rd}$ quartile ($75^{th}$ percentile) and IQR represents the inter-quartile range (i.e. difference between the $75^{th}$ percentile and $25^{th}$ percentile) respectively. A word cloud representation of the specialties of the prescribers across each of the schedule was subsequently generated.

*Prescriber-Prescriber Network Abstraction*: Prescription pattern across each of the scheduled drugs can be modeled as a bipartite graph, **Fig. 1** with patients and prescribers represented by nodes and edges represented by patients seeking drugs across prescriber(s). The present study focuses on one of the projections (PPN) of the bipartite network representing the associations between prescribers, **Fig. 1**. An edge in the PPN represents an indirect association between two seemingly unrelated prescribers as a result of a patient seeking drugs across them. Since the weights of the edges in the PPN are proportional to the number of distinct patients seeking drugs across a pair of prescribers, they can be thought of as representing the strength of the association between the prescribers. Singleton nodes in PPN can arise when patients do not seek drugs across more than one prescriber. Therefore, PPN by very construction can be represented as a weighted undirected graph.

**Table 2.** Description of Schedule II, III, IV drugs used in the study

| Schedule | Drugs | Description |
|---|---|---|
| Schedule II | Oxycodone-acetaminophen | Opioid Analgesic |
| | Oxycodone hcl | Opioid Analgesic |
| | Methadone hcl | Opioid Analgesic |
| | Morphine sulfate | Opioid Analgesic |
| | Hydromorphone hcl | Opioid Analgesic |
| | Oxymorphone hcl | Opioid Analgesic |
| Schedule III | Acetaminophen-codeine | Opioid Analgesic |
| | Buprenorphine | Opioid Analgesic |
| | Buprenorphine-naloxone | Opioid Analgesic/partial agonist |
| | Butalbital codeine | Opioid Analgesic |
| | Hydrocodone-acetaminophen | Opioid Analgesic |
| Schedule IV | Alprazolam | Benzodiazepine |
| | Carisoprodol | Skeletal Muscle Relaxant |
| | Chlordiazepoxide hcl | Benzodiazepine |
| | Clonazepam | Benzodiazepine |
| | Clorazepate dipotassium | Benzodiazepine |
| | Diazepam | Benzodiazepine |
| | Flurazepam hcl | Benzodiazepine |
| | Lorazepam | Benzodiazepine |
| | Midazolam hcl | Benzodiazepine |
| | Oxazepam | Benzodiazepine |
| | Pentazocine-naloxone hcl | Opioid Analgesic/partial agonist |
| | Phenobarbital | Barbiturates |
| | Temazepam | Benzodiazepine |
| | Triazolam | Benzodiazepine |
| | Zolpidem tartrate | Benzodiazepine |

*Surrogate testing*: Surrogate testing [21-24] is a resampling technique for hypothesis testing using a single empirical sample. Essential ingredients of surrogate testing include (*a*) null hypothesis (*b*) surrogate algorithm and (*c*) discriminant statistic. The surrogate algorithm generates independent constrained randomized realizations (surrogates) from the given empirical sample under implicit constraints (i.e. constraints). The constraints are closely related to the null hypothesis of interest and retain certain critical properties of the empirical sample in the surrogates. The discriminant statistic is chosen such that its estimate on the surrogate realizations and the empirical samples is significantly different when the null hypothesis is rejected. In the present study, parametric testing was used to assess the significance of surrogate testing. By definition, parametric testing rejects the null hypothesis if

$$S = \frac{|m_{orig} - \mu_{surr}|}{\sigma_{surr}} > 2,$$

In the above expression, $m_{orig}$ represent the discriminant statistic obtained on the empirical sample whereas $(\mu_{surr}, \sigma_{surr})$ represent the mean and standard deviation of the discriminant statistic estimates obtained on the $n_s$ independent surrogate realizations. The number of surrogate realizations was fixed at $n_s = 99$ [23,24] in the present study.

*Community Structures in PPN*: Several community structure detection algorithms [25,26] have been proposed in the literature to generate a natural partitioning of a given network. These broadly fall under overlapping and non-overlapping community structure detection algorithms [27]. Since a prescriber can be a member of more than one community, PPN communities were investigated using overlapping community structure detection [26]. Subsequently, surrogate testing was used to test whether the community-structure of the PPN were similar to those generated by synthetic random graph models [21,22]. Rather than consider arbitrary random graph models, we considered random graphs that retains the degree centrality distribution [28] of the PPN. Retaining the degree distribution also retains the number of nodes, edges and possibly other critical characteristics (e.g. scale-invariance) of the PPN in the surrogate realization. Partition density [26] measures the quality of link partition in a given network. Therefore, maximum partition density was used as the discriminant statistic in the surrogate testing procedure. The maximum partition density estimation determines the similarity between the edges in the PPNs using Jaccard coefficient. Subsequently, the edges are hierarchically clustered using average linkage. The resulting dendrogram is cut at the point that maximizes the density of edges within the clusters normalized by the maximum and minimum number of edges possible in each cluster. A formal treatment of the community structure detection approach and its implementation details can be found elsewhere[26,29].

**Table 3.** Number of nodes and edges in the PPN and its largest connected component

| Scheduled Drugs | PPN | | Largest Connected Component of PPN | |
|---|---|---|---|---|
| | Nodes | Edges | Nodes | Edges |
| Schedule II | 2764 | 683 | 113 | 136 |
| Schedule III | 4862 | 2654 | 1613 | 2229 |
| Schedule IV | 3970 | 1376 | 879 | 1035 |

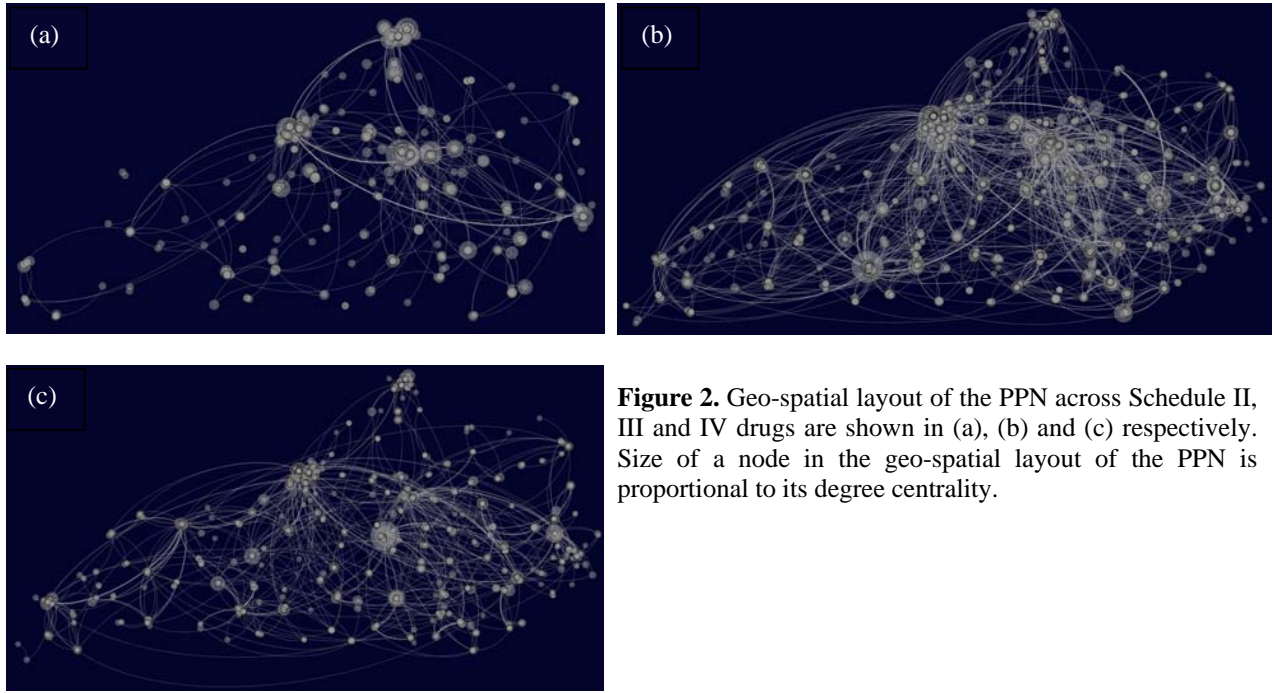

**Figure 2.** Geo-spatial layout of the PPN across Schedule II, III and IV drugs are shown in (a), (b) and (c) respectively. Size of a node in the geo-spatial layout of the PPN is proportional to its degree centrality.

## Results

*Visualization of the PPNs*: The number of nodes for Schedule II, III, IV drugs (**Table 3**) includes singleton nodes, isolated clusters as well as the giant connected component. The giant connected component represents the largest subgraph where it is possible to traverse between any pair of nodes. Geo-spatial layout of the PPN after eliminating all the singleton nodes for Schedule II, III, IV generated using the open-source network visualization tool Gephi 0.8.2[30] is shown in **Figs. 2a-2c** respectively. Latitude and longitudinal coordinates were available across a majority of the prescribers of Schedule II (~93%), Schedule III (~94%) and Schedule IV (~93%). Geo-spatial layout using provides a convenient visualization of patient-movement across multiple prescribers and geographically distinct regions[12,31] as well as the non-uniform spatial distribution of the edges and pockets of high-activity in the state across the three schedules. Ideally, such intricate patient movement is not expected for prescription data across a single month. Degree centrality distribution of the PPNs for Schedule II, III and IV drugs were positively skewed with a small proportion of highly connected prescribers comprising the tails of these distributions. The skewness ($\psi$) and kurtosis ($\kappa$) were Schedule II ($\psi \sim 3, \kappa \sim 13$), Schedule III ($\psi \sim 3.5, \kappa \sim 21$) and Schedule IV ($\psi \sim 6, \kappa \sim 71$) followed the ranking (Schedule IV > Schedule III > Schedule II) across each of these measures. Such skewed distribution may also challenge random graph models such as Erdos-Renyi graphs[32] as potential generative mechanism of PPN. It is important to note that the skewed degree centrality distribution is not necessarily related to non-uniform distribution of the patients across the state. While the former reflects the patient seeking scheduled drugs across multiple prescribers in the state possibly across geographically diverse locations, the latter represents variation in the number of patients in each local region in the state. The markedly skewed degree centrality distributions may also reflect inherent robustness of the PPNs to random interventions as well as susceptibility to targeted interventions of the highly connected nodes. More specifically, targeted interventions of the highly connected prescribers can have the potential to considerably alter the PPN topology and disrupt the patient-movement patterns.

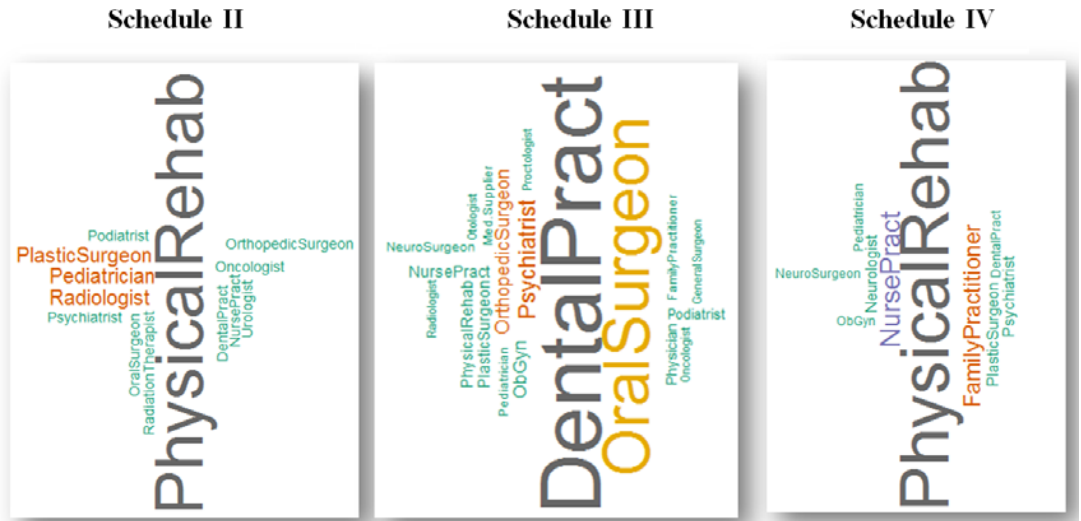

**Figure 3.** Word cloud representing a qualitative description of the specialties of the prescribers with questionable patterns across Schedule II, III and IV drugs using the outlier detection approach. The size of the font is proportional to the frequency of occurrence.

*Prescribers with Questionable Patterns*: Using the outlier detection approach with recommended threshold[20], the number of prescribers flagged across Schedule II, Schedule III and Schedule IV drugs were 77, 187, and 110 respectively. A word cloud representation of the frequency of the specialties across the flagged prescribers is shown in **Fig. 3**. Specialties such as General Practitioners and Internist were to be expected and consistently appeared as the most prevalent across each of the schedules. These were removed and the word cloud was regenerated. Schedule III was dominated by Dentistry whereas those of Schedule II and IV were dominated by Physical Rehabilitation. Specialties that were consistently expressed across the three schedules consisted of General Practitioner, Internist, General Pediatrician, Physical Medicine and Rehabilitation Practitioner, Psychiatrist, Nurse Practitioner and Plastic Surgeon.

*Community Structure Detection*: Since the community structure detection required the network to be connected we considered the largest connected component of the PPN across each of the Scheduled drugs, **Table 3**. Maximum partition density in conjunction with surrogate testing resulted in sigma estimates across Schedule II (S ~ 15 >> 2), Schedule III (S ~ 86 >> 2) and Schedule IV (S ~ 36 >> 2) rejecting the null hypothesis that the community structures of the PPNs corresponding to scheduled drugs were similar to those that can be generated by synthetic random graph models that retain the degree centrality distribution of the PPN.

**Conclusion**

Prescription drug abuse has been identified as a major public health crisis in the United States. Several states have created prescription drug monitoring programs that collect information on controlled substance along with patient and prescriber information. State regulatory agencies use these data sets to identify high volume prescribers or high use patients and may refer cases to licensing board investigators or law enforcement in an effort to discourage doctor shopping, inappropriate prescribing, drug diversion, and inappropriate drug dispensing[33].

The present study investigated the concerted working of prescribers as a system using network abstractions (PPN) across Schedule II, III, IV drugs in a Kentucky Medicaid population. More specifically, such system-level representations can reveal characteristics not readily apparent from classical reductionist representations that treat prescribers as independent entities. The investigation was restricted to prescription data in a single month with minimal chances of patients seeking drugs within a given schedule across multiple prescribers. However, PPNs across Schedule II, III, IV drugs revealed intricate topologies, positively skewed degree centrality distributions and non-trivial community structures markedly different from those of synthetic random graph models. These results reflect the inherent resilience of PPNs to arbitrary intervention and encourage targeted monitoring and surveillance of select prescriber communities in the PPN. Geospatial layout of the PPNs also revealed patient movement across large distances. While the average prescription count was positively skewed, univariate outlier detection using recommended and conservative thresholds revealed specialties of prescribers that were constitutively flagged across Schedule II, III and IV drugs.

The S estimates of the PPN from surrogate testing across Schedule III was relatively higher potentially reflecting more intricate connectivity pattern followed by Schedule IV and Schedule II in that order. Considering the different drugs in each schedule, we expect Schedule II to have the greatest abuse potential compared to Schedule III and IV. A possible explanation could be that prescribers are more reluctant to prescribe Schedule II drugs given the greater danger for abuse and overdose potential. In addition, there are regulatory and workflow differences for Schedule II compared to other schedules. For instance, Schedule II prescriptions undergo greater scrutiny from regulators and do not use refills while all other prescriptions may have up to five refills every six months requiring frequent patient visits. In contrast, Schedule III drug Hydrocodone-Acetaminophen, is the most prescribed drug in the United States. It is used for short term pain relief for multiple conditions such as dental surgery, injuries, and minor surgery. More recently, FDA moved hydrocodone and hydrocodone combination medications from Schedule III to Schedule II in 2014 in an effort to minimize its overuse and abuse. The primary drugs in Schedule IV are benzodiazepines, used for anxiety, but may also be abused in combination with an opioid pain reliever such as Hydrocodone or oxycodone. While the results were presented on a cross-sectional prescription data in a single month in a Medicaid population, the approach can be extended to investigate temporal evolution of prescription patterns as well as changes in prescription patterns in a pre-/post- setting. The S estimates from surrogate testing essentially quantifies the extent of deviation of the PPN patterns from random graph surrogates and has the potential to serve as a useful metric in evaluation of the PPN topology in response to intervention. For instance, targeted intervention of select communities has the potential to considerably fragment the PPNs resulting in markedly lower S estimates. It might not be uncommon to encounter scenarios that are accompanied by a significant drop in the S estimates upon targeted intervention. For instance, a PPN with intricate patterns (S >> 2) upon successful intervention may result in S estimates (S << 2) comparable to those of the random graph surrogate counterparts.

There are several limitations with the present study. In the present study, data with missing values across any of the attributes of interest were excluded from further consideration. While the PPNs were modeled as undirected graphs with edges having unit weight, a more realistic representation of PPN would be weighted undirected graph where the weights correspond to the number of patients/duplicate visits that are shared by a pair of prescribers. While statistically motivated approaches based solely on the distribution of the prescription counts were used to flag questionable prescription patterns in the present study, it might be useful to incorporate the topological properties of the PPNs in the flagging procedure. This can be accomplished by treating all the attributes of interest as features within a binary classification framework. Such an approach obviates the need for pre-defined thresholds and instead relay on the classifier decision boundary given by the optimal combination of the features of interest in discerning questionable prescribers from others. However, such an analysis implicitly demands ground truth class labels and the small number of prosecuted prescribers can contribute to class imbalance issues. While the present study rejected a particular class of random graph surrogates as potential generative mechanism of the communities in PPNs, it did not identify the mechanism that generates PPNs. The present study also restricted the analysis to a single cross-sectional data from the Kentucky Medicaid population. A more detailed investigation of PPN properties across multiple states may be necessary in order to identify universal characteristics, metrics and establish the generalizability of findings. Such an analysis would demand access to the prescription data sets across state boundaries and may be an important step in realizing an interactive national geo-spatial map for real-time monitoring of prescription patterns.